\documentclass[aps,prb,twocolumn,superscriptaddress,showpacs,amsmath,amsfonts,
               floatfix]{revtex4}

\usepackage{epsfig,psfrag}
\newcommand{\ket}[1]{\left|#1\right\rangle}
\newcommand{\bra}[1]{\left\langle#1\right|}
\newcommand{\erw}[1]{\left\langle#1\right\rangle}

\begin{document} 

\title{Dicke Effect in the Tunnel Current through two Double Quantum Dots}

\author{T. Vorrath}
\affiliation{Universit\"at Hamburg, I.~Institut f\"ur Theoretische Physik, 
             Jungiusstr.~9, 20355 Hamburg, Germany}
\author{T. Brandes}
\affiliation{Department of Physics, University of Manchester, 
             UMIST, Manchester M60 1QD, UK}

\date{\today}

\begin{abstract}
We calculate the stationary current through two double quantum dots which
are interacting via a common phonon environment.
Numerical and analytical solutions of a master equation in the stationary 
limit show that the current can be increased as well as decreased due to a 
dissipation mediated interaction. This effect is closely related to collective,
spontaneous emission of phonons (Dicke super- and subradiance effect), and 
the generation of a `cross-coherence' with entanglement of charges in singlet 
or triplet states between the dots. Furthermore, we discuss an inelastic 
`current switch' mechanism by which one double dot controls the current of 
the other.
\end{abstract}

\pacs{
73.21.La,     
73.63.Kv,     
85.35.Gv,     
03.65.Yz      
}
\maketitle

\section{Introduction}
The interaction with a dissipative environment can considerably modify
the physics of very small systems which are described by a few quantum
mechanical states only. \cite{Legetal87}
One may think of an excited atom that decays
via the emission of a photon due to the coupling to the radiation field.
\cite{Allen}
The influence of the environment becomes even more important if not only 
a single system is coupled to it but many. This 
introduces an indirect interaction between the otherwise 
independent systems which can result in an entanglement and collective effects
of the small systems. In the case of identical excited atoms, the interaction
to the common radiation field strongly affects the emission 
characteristics and leads to a collective spontaneous emission, the
so-called superradiance, as first pointed out by R.~H.~Dicke
\cite{Dic54,Andreev,Benedict}
nearly half a century ago.

The influence of a dissipative environment on a single two-level system,
the smallest non-trivial quantum system, has been studied
extensively with the spin-boson-model \cite{Legetal87}
where the environment is modeled by a continuum of harmonic oscillators.
Especially useful for the experimental realization of two level systems are 
coupled semiconductor quantum dots
as these allow tuning of the parameters over a wide range.
\cite{Vaartetal95,Blietal96,Fujetal98,Blietal98b,Taretal99}
Moreover, in these systems
transport spectroscopy is possible by connection with leads
\cite{Fujetal98,Taretal99,SW96,SN96,GP96,Gur98,BBS97,Oosetal98,Blietal98a,
      SWL00,Holetal00,BR00,BRB01,BV02}.
The dissipative environment  
is given by the phonons of the sample and governs
the inelastic current through the system.
\cite{Fujetal98,Taretal99,Brandesddot,Qinetal01,BV01,Fujetal02,DBK02,Hoeetal02}
The electron spin \cite{LD98,EL01,PSL02,SL02} or the electron charge 
\cite{BL00,BV02} in quantum dots have also
been suggested to provide a controllable
realization of scalable qubits. \cite{LD98,CBL00,EL01,CB01,MSS01,PL02}
Arrays of double quantum dots \cite{ZR98}
correspond to charge qubit `registers', and simple `toy' models of $N$ coupled
two-level systems have been used to study collective decoherence effects
in qubit registers.\cite{PSE96,RQJ02,YE02}
Furthermore, controllable two-level systems with Cooper pairs tunneling 
to and from a superconducting box have been realized experimentally. 
\cite{NPT99,Vioetal02}

Coherent effects in small clusters of 
two level systems caused by the coupling to a common
environment have been realized mainly in the field of quantum optics.
In ion laser traps, Dicke sub- and superradiance has been measured 
by DeVoe and Brewer \cite{DeVB96} in the spontaneous emission rate of photons 
from two ions as a function of the ion-ion distance.
Furthermore, entanglement in linear ion-traps can be generated by the coupling
of (few-level) ions to a common single bosonic mode, the center-of-mass 
oscillator (vibration) mode \cite{MS99,SM99,Sacetal00}.
Even the generation of entangled light from white noise \cite{PH02} has been 
suggested. 

The appearance of collective quantum optical effects in mesoscopic 
transport has re-gained considerable interest quite recently.  
Shahbazyan and Raikh \cite{SR94} first predicted the Dicke (spectral function)
effect \cite{Dic53}  to appear in resonant tunneling through two impurities, 
which was later generalized to scattering properties in a strong magnetic field
\cite{SU98}. The Dicke effect was predicted theoretically in `pumped', 
transient superradiance of quantum dot arrays coupled to electron reservoirs 
\cite{BIS98}, and in the AC conductivity of dirty multi-channel quantum wires 
in a strong magnetic field. \cite{Bra00} 

In this work, we focus on coherent effects in mesoscopic few level 
systems. As a realization, we choose two nearby but otherwise independent 
double quantum dots coupled to the same phonon environment. 
We study the influence of the resulting indirect interaction
on the transport properties and calculate the stationary current.
Signatures of `super'- and sub-radiance of {\em phonons} are predicted
which show up as an increase or a decrease of the stationary electron current. 
We demonstrate that this effect is directly related to the creation of 
{\em charge} wave function entanglement between the two double dots, which 
appears  in a preferred formation of either a (charge) triplet or singlet 
configuration, depending on the internal level splittings and/or the tunnel 
couplings to the external electron leads in both sub-systems.
Generation of entanglement via phonons
becomes attractive in the light of recent investigations of 
single-electron tunneling through individual molecules 
\cite{Paretal00,JGA00,Reietal02,BS01,GK02}, or quantum dots in freestanding 
\cite{SCCR95,CR98,BRWB98,Blietal00} 
and movable \cite{Goretal98,WZ99,AWZB01,AM02} nano-structures,
in both of which vibration properties on the nanoscale 
seem to play a big role.

The outline of the paper is as follows. In section \ref{sectionmodel} we 
introduce the model and our method. Current superradiance is discussed in 
section \ref{sectionsuper}. Section  \ref{sectionsub} presents current 
subradiance and the inelastic current `switch' mechanism. Finally, we conclude
in section \ref{sectionconclusion}.

\section{Model and method}\label{sectionmodel}
\begin{figure}
\centering
\psfrag{Gl}{$\Gamma_L$}
\psfrag{Gr}{$\Gamma_R$}
\psfrag{Tc}{$T_c$}
\psfrag{e1}{$\varepsilon_L$}
\psfrag{e2}{$\varepsilon_R$}
\psfrag{Gl1}{$\Gamma_{L,1}$}
\psfrag{Gr1}{$\Gamma_{R,1}$}
\psfrag{Tc1}{$T_{c,1}$}
\psfrag{Gl2}{$\Gamma_{L,2}$}
\psfrag{Gr2}{$\Gamma_{R,2}$}
\psfrag{Tc2}{$T_{c,2}$}
\epsfig{file=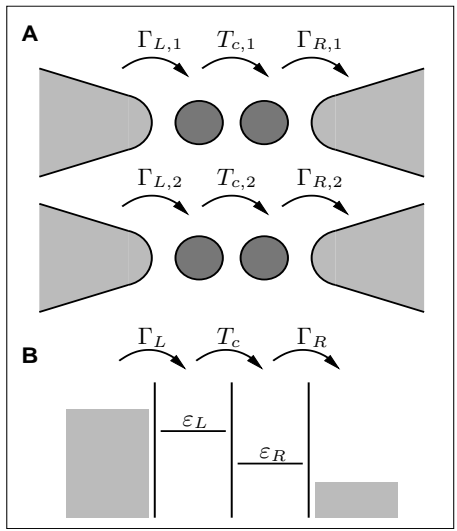,width=6cm,height=7cm}
\caption{A: $N\!=\!2$ `charge qubit register' with two double quantum dots 
coupled to independent electron leads. B: energy diagram of one individual 
double dot.}
\label{dots}
\end{figure}

Our model is a system (`register') of two double quantum dots (DQDs), 
each of which consist of two individual quantum dots (called `left' and 'right'
in the following). Both double dots are coupled to independent left and right
leads as depicted in Fig.~\ref{dots}~A.

We concentrate on boson-mediated collective effects 
between the DQDs originating from the coupling of the whole system 
to a common dissipative, bosonic bath that will be specified below.
In the following we completely neglect static tunnel coupling between
the individual DQDs and, more important, inter-DQD
Coulomb correlations. Although this is a severe limitation for the general
applicability of the model, it still grasps the essential physics of 
dissipation induced entanglement. However, one might envisage configurations 
with intradot Coulomb matrix elements much larger than interdot 
matrix elements.

In this paper, we choose the simplest possible description
of an environment coupling in close analogy to the standard
spin-boson Hamiltonian \cite{Legetal87}.
The results of this model for the tunnel current through 
one double dot are in relatively good agreement with 
experimental observations \cite{Brandesddot,BV02}.
The role of off-diagonal terms in a single DQD has been 
discussed recently \cite{KS_02}.

\subsection{Hamiltonian}
The Hamiltonian and the subsequent derivation of the master equation is
given for the general case of $N$ double quantum dots.
We study the stationary tunnel current through the dots with all lead 
chemical potentials such that electrons can only flow from the left to the 
right. Furthermore, we restrict ourselves to the strong Coulomb blockade 
regime in each individual double dot where only one additional electron is 
allowed on either the left or the right dot. The Hilbert space of the $i$-th 
double dot then is spanned by the three many-body states $\ket{L,i}$ 
(one additional electron in the $i$-th left dot at energy $\varepsilon_{L,i}$),
$\ket{R,i}$ (one additional electron in the $i$-th right dot at energy 
$\varepsilon_{R,i}$), and $\ket{0,i}$ (no additional electron in either of the
dots). For $N=1$ this has been proven \cite{SN96,Brandesddot}
to be a valid description of non-linear transport experiments in double 
quantum dots \cite{Fujetal98,Taretal99}.

Introducing the operators
\begin{align} \nonumber
n_{L,i} &= \ket{L,i} \bra{L,i} & n_{R,i} &= \ket{R,i} \bra{R,i}\\
p_i &=  \ket{L,i} \bra{R,i}    & p_i^{\dagger} &= \ket{R,i}\bra{L,i}\\
s_{L,i} &= \ket{0,i} \bra{L,i} & s_{R,i} &=  \ket{0,i}\bra{R,i},
\nonumber
\end{align}
the total Hamiltonian can be written as
\begin{equation}
\label{Hamiltonian}
\begin{split}
H=&\sum_i^N \bigg[ \varepsilon_{L,i} \, n_{L,i} + \varepsilon_{R,i} \, n_{R,i}
   + T_{c,i} (p_i + p_i^{\dagger}) \\
   &+ \sum_{k} V_{k,i}^L \, c_{k,i}^{\dagger} s_{L,i} + h.c. 
   + \sum_k \varepsilon_k^L \, c_{k,i}^{\dagger} c_{k,i}\\
   &+ \sum_l V_{l,i}^R \, d_{l,i}^{\dagger} s_{R,i} + h.c.
   + \sum_l \varepsilon_l^R \, d_{l,i}^{\dagger} d_{l,i}\\
   &+ \sum_q \gamma_{q,i}\,(a_q^{\dagger} + a_{-q}) (n_{L,i} - n_{R,i}) \bigg]
   + \sum_q \omega_q \, a_q^{\dagger} a_q \, .
\end{split}
\end{equation}
Here, the electrons in mode $k\;(l)$ with energy $\varepsilon_{k(l)}^{L(R)}$ 
in the left (right) leads pertaining to DQD $i$ are described by 
creation operators $c_{k,i}^{\dagger}$ ($d_{l,i}^{\dagger}$), and 
the coupling matrix elements to the leads are denoted by $V_{k,i}^{L/R}$.
A boson in mode $q$ with energy $\omega_q$ is created by the 
operator $a_q^{\dagger}$. As in the standard spin-boson model, 
we assume a simplified coupling to the quantum dots which is purely 
diagonal with matrix element $\gamma_{q,i}$ for mode $q$ to the 
$i$-th double dot.

So far, no further assumptions have been made with respect to 
the specific realization of the DQDs and the dissipative bath.
Nevertheless, the system we have in mind are lateral or vertical double dots, 
where the primary bosonic coupling has been shown due to phonons of the 
semiconductor substrate. The microscopic details determine the tunnel matrix 
elements $T_{c,i}$, $V_{k,i}^{L/R}$, and the electron-phonon coupling 
constants $\gamma_{q,i}$.

\subsection{Density matrix}
In the following, we employ a master equation description for the time 
evolution of the register within the Born-Markov approximation, which takes 
into account the interactions with the leads and the bosonic environment up 
to second order. Alternatively, electron-phonon interactions can be
treated exactly by a polaron transformation \cite{Brandesddot,BV02} and 
perturbatively in the tunnel matrix elements 
$T_{c,i}$. For $T_{c,i}\lesssim |\varepsilon_{Li}-\varepsilon_{Ri}|$
and small coupling to the bosonic bath, the results of both methods 
practically coincide \cite{BV01}.

The time derivative of the reduced density matrix $\rho(t)$ of the double 
quantum dots is given by 
\begin{equation}
\label{born-markov}
\begin{split}
\dot{\tilde{\rho}}(t) = - \int_0^t \! dt'\; 
&\Big( {\rm Tr}_{\,\textrm{Res,e}} \left\{ \big[\tilde{H}_{\rm e}(t),
 \big[ \tilde{H}_{\rm e}(t'),\tilde{\rho}(t) 
   \otimes \tilde{R}_{0,\rm e} \big] \big] \right\} \\
+ &{\rm Tr}_{\,\textrm{Res,\rm p}} \left\{ \big[\tilde{H}_{\rm p}(t),
 \big[ \tilde{H}_{\rm p}(t'),\tilde{\rho}(t) 
   \otimes \tilde{R}_{0,\rm p} \big] \big] \right\} \Big),
\end{split}
\end{equation}
where the tilde indicates the interaction picture, $H_{\rm e}$ ($H_{\rm p}$) 
denotes the interactions between the double dots and the leads (the phonons), 
and $R_{0,\rm e}$ ($R_{0,\rm p}$) is the density matrix of the leads 
(the phonons). Equation~(\ref{born-markov}) is the sum of an electron and 
a phonon part since we neglect correlations between leads and phonons.

The trace over the equilibrium electron reservoirs, 
${\rm Tr}_{\,\textrm{Res,e}}$, results in Fermi functions of the leads.
As we are interested in large source-drain voltages between the left and 
the right leads, the Fermi functions of the left leads can be set to one 
and those of the right leads to zero. Moreover, the energy dependence
of the tunnel rates 
\begin{equation}
 \Gamma_{L/R,i} = 2 \pi 
\sum_{k} |V_{k,i}^{L/R}|^2 \; \delta(\varepsilon - \varepsilon_{k}^{L/R})
\end{equation}
is neglected.

\subsubsection{Electron-phonon interaction}
In the following, we consider identical electron-phonon interaction 
in the DQDs,
\begin{equation}\label{gammai}
\gamma_{q,i} = \gamma_q \,. 
\end{equation}
Depending on the relative position of the quantum dots (lateral, vertical), 
the electron wave functions in the dots, and the geometry of the phonon 
substrate (bulk, slab \cite{DBK02}, sheet etc.), the $\gamma_{q,i}$  will 
never be exactly identical in real situations. Therefore,
Eq.~(\ref{gammai}) can only be regarded as an idealized 
limit of, e.g., a phonon resonator or a  situation where the distance between 
different double dots is small as compared to the relevant phonon wavelengths.

We define a correlation function of the boson system
\begin{equation}
K(t)\equiv 
\int_0^{\infty}\!d\omega \; \rho(\omega)\;  
\frac{e^{i \omega t} e^{-\beta \omega} + e^{-i \omega t}}
{1-e^{-\beta \omega}}\; 
\end{equation}
that results from the trace over the bosonic degrees of freedom.
Here, $\beta=1/k_BT$ denotes the inverse phonon bath temperature, and 
the spectral function $\rho(\omega)$ of the bosonic environment is defined as 
\begin{equation}
\label{spectral_fct}
\rho(\omega) \equiv \sum_q |\gamma_{q}|^2 \; \delta(\omega-\omega_q).
\end{equation}
For the calculations, we use the spectral function of bulk acoustic 
phonons with piezoelectric interaction to electrons in lateral quantum dots 
\cite{Brandesddot,BV01},
\begin{equation}
\label{ac_phonons}
\rho(\omega) = g \,\omega \,\Big(1 - \frac{\omega_d}{\omega} 
            \sin\Big(\frac{\omega}{\omega_d}\Big)\Big)\;e^{-\omega/\omega_c},
\end{equation}
where $g$ is the dimensionless interaction strength, $\omega_c$ the cut-off
frequency and the frequency $\omega_d$ is determined by the ratio of the
the sound velocity to the distance between two quantum dots.

In the following, integrals over $K(t)$ are required as
\begin{equation}\label{GammaCSdefinition}
\begin{split}
\Gamma_{C,i} &\equiv \int_0^{\infty} \!  K(t) \, \cos(\Delta_i t)\;dt 
   = \frac{\pi}{2}\;\rho(\Delta_i)\,
      \coth \!\Big(\frac{\beta \Delta_i}{2}\Big),\\
\Gamma_{S,i} &\equiv \int_0^{\infty} \!  K(t) \, \sin(\Delta_i t)\;dt\,
   = -i \;\frac{\pi}{2}\;\rho(\Delta_i).
\end{split}
\end{equation}
with the hybridization energy $\Delta_i=(\varepsilon_i^2+4 T_{c,i}^2)^{1/2}$ 
and the energy bias
$\varepsilon_i = \varepsilon_{L,i} \!-\! \varepsilon_{R,i}$ in the $i$-th dot.
The integrals are calculated neglecting the principal values \cite{BV02}. We  
furthermore assume a spectral function $\rho(\Delta_i)$ such that
$\Gamma_{C,i} \to 0$ for $\Delta_i \to 0$ which is fulfilled for
microscopic models of the electron-phonon interaction in double quantum dots 
\cite{Brandesddot,BV02}.

\subsubsection{Master equation}
Inserting the traces over the electron reservoirs and the bosonic bath
into Eq.~(\ref{born-markov}) and transforming back to Schr\"odinger picture 
yields a master equation for the reduced density 
matrix of the total DQD register,
\begin{equation}
\label{master_operator}
\begin{split}
&\dot{\rho} (t) 
  = i \; \sum_{i=1}^N \Big\{ \Big[\,\rho(t)\,, \varepsilon_{L,i} \, n_{L,i} 
  + \varepsilon_{R,i} \, n_{R,i} + T_{c,i} (p_i + p_i^{\dagger}) \Big] \\
&+ \frac{\Gamma_{L,i}}{2} \Big(
  2 \, s_{L,i}^{\dagger}\,\rho(t)\,s_{L,i}
  - \,s_{L,i}\,s_{L,i}^{\dagger}\,\rho(t)
  - \,\rho(t)\,s_{L,i}\,s_{L,i}^{\dagger} \Big) \\
&+ \frac{\Gamma_{R,i}}{2} \Big(
  2 \, s_{R,i}\,\rho(t)\,s_{R,i}^{\dagger}
  - s_{R,i}^{\dagger}\,s_{R,i}\,\rho(t)
  - \rho(t)\,s_{R,i}^{\dagger}\,s_{R,i} \Big) \Big\} \\
&-  \sum_{i,j} \, \Big\{ \Big[(n_{L,i}\!-\!n_{R,i}), A_j\, \rho(t) \Big]
          \!-\! \Big[(n_{L,i}\!-\!n_{R,i}), \rho(t) A_j^{\dagger} \Big] \Big\},
\end{split}
\end{equation}
with
\begin{equation}
\begin{split}
A_j &\equiv  \frac{2 T_{c,j}}{\Delta_i^2} \,
      \Big( \,  2T_{c,j} \Gamma_{C,j} (n_{L,j}\!-\!n_{R,j}) \\
    &- \Gamma_{C,j}\, \varepsilon_j \, (p_j + p_j^{\dagger})
      + i\,  \Delta_j \Gamma_{S,j} (p_j - p_j^{\dagger})\, \Big).
\end{split}
\end{equation}
From Eq.~(\ref{GammaCSdefinition}) it is obvious 
that the influence of the bosonic bath enters only via the spectral 
functions $\rho(\omega)$
as defined in Eq.~(\ref{spectral_fct}). All microscopic properties of the 
phonons and their interaction mechanism to the electrons in the quantum dots
are described by these functions.

Furthermore, we point out that the mixed terms $i\not=j$ in 
Eq.~(\ref{master_operator}) are responsible for the collective effects 
to be discussed in the following. Without these terms,
the master equation would merely describe an ensemble of $N$ 
independent DQDs. In that case, an initially 
factorized density matrix of the total system would always remain 
factorized and no correlations could build up.
The terms $i\not=j$ introduce correlations between the different
double dots, the origin of which lies in the coupling to the same bosonic 
environment.

\section{Current superradiance}\label{sectionsuper}
We restrict ourselves to the stationary case where the time derivative
of the density matrix, $\dot{\rho}(t)$, vanishes. 
Then, Eq.~(\ref{master_operator}) reduces to a linear system of equations 
which can be easily solved numerically.
Results for a single double quantum dot, $N\!=\!1$, can be 
obtained analytically \cite{BV01} and are given for two expectation values
below, Eq.~(\ref{solution}). For $N>1$, the dimension of 
the density matrix grows as $9^N$ (although not all of the matrix elements
are required) whence analytical solutions become very cumbersome.
For the rest of this paper, we restrict ourselves to the case of two 
double dots ($N=2$), called DQD 1 and DQD 2 in the following.

\subsection{Stationary current}
The total electron current is simply given by the sum of the 
currents through the individual DQDs, as electrons cannot tunnel between
different double dots. The current operator of DQD $i$ is
\begin{equation}
I_i = \frac{i \,T_{c,i} \,e}{\hbar}\; (p_i-p_i^{\dagger}),
\end{equation}
and the corresponding expectation values 
can be expressed by the elements of the density matrix as
\begin{equation}
\label{current}
\begin{split}
I_1 &= - \,\frac{2 \,T_{c,1}\, e}{\hbar} \; \;
\mbox{Im} \big\{ \rho_{LRLL} + \rho_{RRLR} + \rho_{0RL0} \big\},\\
I_2 &= - \,\frac{2 \,T_{c,2}\, e}{\hbar} \; \;
\mbox{Im} \big\{ \rho_{RLLL} + \rho_{RRRL} + \rho_{R00L} \big\},
\end{split}
\end{equation}
with the notation
\begin{equation}
\rho_{j\,i\,i'\,j'} = \;
 {}_2 \!\bra{j} \otimes {}_1 \!\bra{i}\,\rho \,\ket{i'}_1 \otimes \ket{j'}_2,
 \quad i,j \in \{ L, R, 0\}.
\end{equation}
The set of linear equations corresponding to Eq.~(\ref{master_operator}) for
$N\!=\!2$ is given in appendix~\ref{appendix_A}, Eq.~(\ref{master_elements}).
\begin{figure}
\centering
\psfrag{I12}{\hspace*{-8mm}$(I_1+I_2)/$pA}
\psfrag{e1}{\hspace*{-5mm}$\varepsilon_1/\mu$eV}
\psfrag{eps30}{$\varepsilon_2=30\mu$eV}
\psfrag{eps55}{$\varepsilon_2=55\mu$eV}
\psfrag{eps100}{$\varepsilon_2=100\mu$eV}
\epsfig{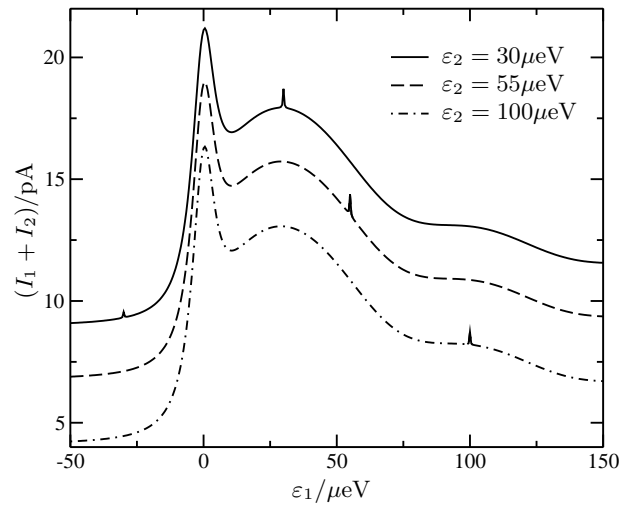}
\caption{
Total current through two double quantum dots as a function of the bias
$\varepsilon_1$. 
The parameters are $T_{c,1}\!=\!T_{c,2}\!=\!3\mu$eV, 
$\Gamma_{L,1}\!=\!\Gamma_{R,1}\!=\!
 \Gamma_{L,2}\!=\!\Gamma_{R,2}\!=\!0.15\mu$eV, and for the spectral function
$g\!=\!0.01$, $T\!=\!23$mK, 
$\omega_d\!=\!10\mu$eV and $\omega_c\!=\!1$meV.
These values are used throughout the whole article if not stated otherwise.}
\label{fI}
\end{figure}
From the numerical solution of Eq.~(\ref{master_elements}) we 
find the stationary current through two double quantum dots as a function
of the bias $\varepsilon_1$ in the first double dot while the bias 
$\varepsilon_2$ in the second is kept constant, cf. Fig.~\ref{fI}.
The overall shape of the current is very similar to the case of one individual
double quantum dot \cite{Brandesddot,BV01}, with its strong elastic peak 
around $\varepsilon_1=0$ and a broad inelastic shoulder for
$\varepsilon_1>0$. The interesting new feature here is 
the peak at the resonance $\varepsilon_1\!=\!\varepsilon_2$ which is 
due to collective effects to be analyzed now.

\subsection{Cross coherences}
\label{subsec_analytisch}
The effective interaction between the two DQDs results from the 
simultaneous coupling of both double dots to the same phonon environment.
It appears in the master equation~(\ref{master_operator})
as the mixed terms $i\!\neq\!j$ in the sum. In the explicit form
of the master equation~(\ref{master_elements}), the effective interaction
is connected to six matrix elements only (and their complex conjugates). 
These elements are 
$\rho_{RLLL}$, $\rho_{LRLL}$, $\rho_{RRLR}$ and $\rho_{RRRL}$,
all of which enter the expression for the current, Eq.~(\ref{current}),
and the two `cross coherence' matrix elements
\begin{equation}\label{crossdefinition}
\rho_{RLRL} = \big\langle p_1^{\dagger} p_2 \big\rangle,\quad 
\rho_{RRLL} = \big\langle p_1 p_2 \big\rangle.
\end{equation}
Therefore, we approximate the collective effects caused by the effective
interaction starting from the solution of the non-interacting master equation,
without the mixed terms $i\!\neq\!j$, and assume that only those matrix
elements mentioned above are affected by the interaction.

\begin{figure}
\centering
\psfrag{eps1}{\hspace*{-3mm}$\varepsilon_1/\mu$eV}
\psfrag{RHO}{}
\psfrag{rho6_large}
       {$\textrm{Re}\big\{\big\langle p_1^{\dagger}p_2\big\rangle\big\}$}
\psfrag{rho14_large}{$\textrm{Re}\big\{\big\langle p_1 p_2\big\rangle\big\}$}
\psfrag{eps}{\hspace*{-4mm}$\varepsilon_1/\mu$eV}
\psfrag{rho6}{\hspace*{-8mm}
   $\textrm{Re}\big\{\big\langle p_1^{\dagger}p_2\big\rangle\big\}$}
\epsfig{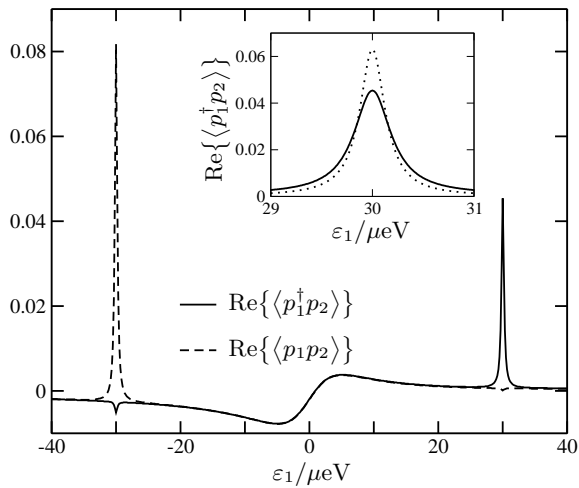}
\caption{Real parts of the cross coherences from
the master equation~(\ref{master_operator}) as functions of the bias 
in the first double dot ($\varepsilon_2\!=\!30 \mu$eV and
the other parameters agree with Fig.~\ref{fI}).
The inset compares the approximation for 
$\textrm{Re}\big\{\big\langle p_1^{\dagger}p_2\big\rangle\big\}$, 
Eq.~(\ref{rho6}) (dotted line), 
with the solution of the master eq. (solid line).
}
\label{F_rho6}
\end{figure}

In the non-interacting case, the cross coherence is simply the product
of the corresponding matrix elements of independent double dots, 
\begin{equation}
\big\langle p_1^{\dagger} p_2 \big\rangle 
= \big\langle p_1^{\dagger} \big\rangle \erw{p_2},  \quad
\big\langle p_1 p_2 \big\rangle = \erw{p_1} \erw{p_2}.
\end{equation}
These can be solved analytically,
\begin{equation}
\label{solution}
\begin{split}
\erw{p_j} &= -\frac{\Gamma_{L,j}}{M_j} 
 \Big( 2 T_{c,j}^2(\beta_j+\gamma_j) \\
   &+ \Gamma_{R,j}(i T_{c,j}+\gamma_j)
     (i \varepsilon_j+\frac{1}{2} \Gamma_{R,j} + 2 \alpha_j) \Big), \\
\erw{n_{L,j}} &= 1- \frac{T_{c,j}(\Gamma_{L,j}\!+\!\Gamma_{R,j})}{M_j}
  \Big( 2 \varepsilon_j \gamma_j + T_{c,j} (\Gamma_{R,j}+4\alpha_j) \Big),
\end{split}
\end{equation}
where $\erw{n_{L,j}}$ is given for later reference, $\alpha_j$, $\beta_j$, 
and $\gamma_j$ as defined in the appendix, Eq.~(\ref{rate_abc}), and with
\begin{equation}
\begin{split}
&M_j \equiv \Gamma_{L,j}\Gamma_{R,j}
 \Big(\varepsilon_j^2+\Big(\frac{1}{2} \Gamma_{R,j}\!+\!2\alpha_j\Big)^2\,\Big)
 - 2 T_{c,j} \varepsilon_j \beta_j \Gamma_{L,j}\\
 &+ T_{c,j}^2 (2\Gamma_{L,j}\!+\!\Gamma_{R,j})(\Gamma_{R,j}\!+\!4\alpha_j)
 + 2 T_{c,j} \varepsilon_j \gamma_j (\Gamma_{L,j}\!+\!\Gamma_{R,j}).
\end{split}
\end{equation}
In the inelastic regime, $T_c\!\ll\!\varepsilon$,
of the non-interacting case, the 
cross coherences $\big\langle p_1^{\dagger} p_2 \big\rangle$ and
$\big\langle p_1 p_2 \big\rangle$ tend to zero as can be seen from
Eq.~(\ref{solution}).
Moreover, we neglect the imaginary part of the cross coherences in the 
interacting case. Then, the change in the current through DQD 1 
due to collective effects can be approximated by
\begin{equation}
\label{Delta_current}
\Delta I_1 = \frac{2 e \,T_{c,1}\, \gamma_2}{\hbar \,\varepsilon_1}\,
 \Big( \textrm{Re}\Big\{ \big\langle p_1^{\dagger} p_2 \big\rangle \Big\}
  -\textrm{Re}\Big\{ \big\langle p_1 p_2 \big\rangle \Big\} \Big).
\end{equation}
Correspondingly, the change $\Delta I_2$ of the current through the 
second double dot DQD 2
is obtained from $\Delta I_1$ by exchanging the subscripts 1 and 2.
Hence, the alteration in the current is proportional to the real parts of
the cross coherences $\big\langle p_1^{\dagger} p_2 \big\rangle$ and 
$\big\langle p_1 p_2 \big\rangle$ between the two DQDs, which
confirms the collective character of the effect.
This result is corroborated by plotting the real parts of 
the cross coherences as a function of $\varepsilon_1$, cf. Fig.~\ref{F_rho6}.
One recognizes that $\big\langle p_1^{\dagger} p_2 \big\rangle$
is peaked around $\varepsilon_1\!=\!\varepsilon_2$,
whereas $\big\langle p_1 p_2 \big\rangle$ has a peak at 
$\varepsilon_1\!=\!-\varepsilon_2$.
The increase of the current at $\varepsilon_1\!=\!\varepsilon_2$ 
is therefore due to the maximum of 
the first correlation $\big\langle p_1^{\dagger} p_2 \big\rangle$.

If we neglect the changes of all other elements of the density matrix
that are caused by the effective interaction between the two DQDs,
the real part of the cross coherence
 $\big\langle p_1^{\dagger} p_2 \big\rangle$
can be approximated around the resonance 
$\varepsilon_1\!=\!\varepsilon_2$ as
\begin{equation}
\label{rho6}
\begin{split}
\textrm{Re}\Big\{\big\langle p_1^{\dagger} p_2 \big\rangle \Big\}
&= \frac{\frac{1}{2} \, (\Gamma_{R,1}+\Gamma_{R,2})}
{(\varepsilon_1\!-\!\varepsilon_2)^2
  +\frac{1}{4}(\Gamma_{R,1}+\Gamma_{R,2})^2} \cdot\\
&\Big( \gamma_1 \, \textrm{Re}\big\{\!\erw{p_2}\!\big\} \erw{n_{L,1}} 
  + \gamma_2 \, \textrm{Re}\big\{\!\erw{p_1}\!\big\} \erw{n_{L,2}} \Big).
\end{split}
\end{equation}
One recognizes that $\big\langle p_1^{\dagger} p_2 \big\rangle$
is Lorentzian shaped as a function of the energy difference 
$\varepsilon_1\!-\!\varepsilon_2$.
The result of Eq.~(\ref{rho6}) with $\erw{p_j}$ and $\erw{n_{L,j}}$
as given in Eq.~(\ref{solution})
is in good agreement with the numerical solution of the
master equation (\ref{master_operator}) (inset of Fig.~\ref{F_rho6}).

Next, we insert the result for the cross coherence 
in Eq.~(\ref{Delta_current}) and find for the change of the tunnel current
due to interaction effects between the two double quantum dots
around the resonance $\varepsilon_1\!=\!\varepsilon_2$:
\begin{equation}
\label{Delta_I1}
\begin{split}
\Delta I_1 = \, &\frac{e \,T_{c,1}\, \gamma_2}{\hbar \,\varepsilon_1}\,
\frac{\Gamma_{R,1}\!+\!\Gamma_{R,2}}
{(\varepsilon_1\!-\!\varepsilon_2)^2
      +\frac{1}{4}(\Gamma_{R,1}\!+\!\Gamma_{R,2})^2} \cdot \\
&\Big( \gamma_1 \, \textrm{Re}\big\{\!\erw{p_2}\!\big\} \erw{n_{L,1}} 
  + \gamma_2 \, \textrm{Re}\big\{\!\erw{p_1}\!\big\} \erw{n_{L,2}} \Big).
\end{split}
\end{equation}
Again, the change in the current through the second double dot, $\Delta I_2$,
is obtained by exchanging the subscripts.
This approximation overestimates the actual change in the current
for the parameters chosen in the previous section
but provides a good qualitative description for the effect of the enhanced 
tunnel current. A comparison between this result and the numerical solution
is given below.

\subsection{Singlet and triplet states}
The collective effects in the two double quantum dots are connected with 
the cross coherence function $\big\langle p_1^{\dagger} p_2 \big\rangle$, 
Eq.~(\ref{crossdefinition}). For the `two-qubit register' one 
can easily prove the operator identity
\begin{equation} \label{PTS}
  p_1^{\dagger} p_2 +  p_2^{\dagger} p_1 = \textrm{P}_{T_0} - \textrm{P}_{S_0},
\end{equation}
where 
$\textrm{P}_{\psi}$ is the projection operator on the state $\ket{\psi}$,
$\textrm{P}_{\psi} \equiv \ket{\psi}\bra{\psi}$, and triplet and singlet 
do not refer to the real electron spin but to the `pseudo' spin
defined in the two dimensional Hilbert space 
${\rm span}(|L\rangle, |R\rangle)$,
\begin{equation}
\label{basis_ts}
\begin{split}
\ket{T_+} &= \ket{L}_1 \ket{L}_2, \quad \ket{T_-} = \ket{R}_1  \ket{R}_2,\\
\ket{T_0} &= \frac{1}{\sqrt{2}} \,
      \Big(\ket{L}_1  \ket{R}_2 + \ket{R}_1 \ket{L}_2 \Big), \\
\ket{S_0} &= \frac{1}{\sqrt{2}} \,
      \Big(\ket{L}_1  \ket{R}_2 - \ket{R}_1 \ket{L}_2 \Big).
\end{split}
\end{equation}
With $2 \, \textrm{Re} \big\langle p_1^{\dagger} p_2 \big\rangle = \big\langle
 \textrm{P}_{T_0} \big\rangle - \big\langle \textrm{P}_{S_0}\big\rangle$ 
and the proportionality $\Delta I_1 \propto \textrm{Re} 
\big\langle p_1^{\dagger} p_2 \big\rangle$ 
for $\varepsilon_1 \approx \varepsilon_2$, cf. Eq.~(\ref{Delta_current}), 
it follows that the current enhancement $\Delta I_1$ is due to an increased 
probability of finding the two electrons in a (pseudo) triplet rather than 
in a (pseudo) singlet state. In the following, we demonstrate that the 
mechanism underlying this effect is indeed the Dicke superradiance effect 
known from quantum optics.

\subsection{Dicke effect}
Superradiance emerges in the collective spontaneous emission from an
ensemble of identical two-level atoms. If $N$ excited atoms are concentrated
in a region smaller than the wavelength of the emitted radiation, they do
not decay independently anymore. Instead, the radiation has a higher
intensity and takes place in a shorter time interval than for an ensemble of
independent atoms due to the coupling of all atoms to the common radiation 
field.

Let us now consider the case $N\!=\!2$ and calculate (similar to the original 
work of Dicke \cite{Dic54}) the decay rate $\Gamma$ of {two} initially 
excited atoms with dipole moments $\hat{{d}}_1$ and  $\hat{{d}}_2$) at 
position ${\bf r}_1$ and ${\bf r}_2$ due to the interaction with light,
\begin{equation}\label{ion}
\begin{split}
H_{eph}&=\sum_{{\bf Q}}{{g}}_{{\bf Q}}
         \left(a_{-{\bf Q}} + a_{{\bf Q}}^+\right)\\
  &\times \left[\hat{{d}}_1\exp{i({\bf Qr}_1)}+
         \hat{{d}}_2\exp{i({\bf Qr}_2)}\right] ,
\end{split}
\end{equation}
from which the spontaneous emission rate  of photons with wave vector 
${\bf Q}$ follows (Fermi's Golden Rule),
\begin{equation}
  \label{eq:gammaplusminus}
  \Gamma_{\pm}(Q)\propto\!\sum_{{\bf Q}} |g_{Q}|^2{|1 
  \pm \exp{[i{\bf Q}({\bf r}_2-{\bf r}_1)]}|^2} 
\delta(\omega_0-\omega_Q) ,
\end{equation}
where $Q=\omega_0/c$, $\omega_0$ is the transition frequency between the upper
and lower level, and $c$ denotes the speed of light. The interference of the 
two interaction contributions $\hat{{d}}_1 e^{i({\bf Qr}_1)}$ and 
$\hat{{d}}_2 e^{i({\bf Qr}_2)}$ leads to a {\em splitting} of the spontaneous 
decay into a fast, `superradiant', decay channel ($\Gamma_{+}(Q)$), and a slow,
`subradiant' decay channel ($\Gamma_{-}(Q)$). 
This splitting is called `Dicke-effect'.

Loosely speaking, the two signs $\pm$ correspond to the two different 
relative orientations of the dipole moments of the two atoms. More precisely, 
from the four possible states in the Hilbert space of two two--level systems,
$  {\cal{H}}_2=C^2 \otimes C^2$,
one can form singlet and triplet states according to 
$  |S_0\rangle :=\frac{1}{\sqrt{2}}\left( |\uparrow\downarrow\rangle
-|\downarrow\uparrow\rangle\right)$,
$  |T_+\rangle :=|\uparrow\uparrow\rangle$,
$  |T_0\rangle :=\frac{1}{\sqrt{2}}\left( |\uparrow\downarrow\rangle
+|\downarrow\uparrow\rangle\right)$, 
and
$  |T_{-}\rangle :=|\downarrow\downarrow\rangle$.
The superradiant decay channel occurs via the triplet and the
subradiant decay via the singlet states \cite{Dic54,GH82}. In the extreme 
`Dicke' limit where the second phase factor is close to unity,
$\exp{[i{\bf Q}({\bf r}_2-{\bf r}_1)]}\approx 1$, it follows that
$\Gamma_{-}(Q)=0$ and $\Gamma_{+}(Q)=2\Gamma(Q)$
where $\Gamma(Q)$ is the decay rate of one {\em single} atom. 
This limit is  theoretically achieved if
$|{\bf Q}({\bf r}_2-{\bf r}_1)|\ll 1$ for all wave vectors ${\bf Q}$, 
i.e. the distance between the two atoms
is much smaller than the wave length of the light. 

We mention that in practice, this `pure' limit, where the subradiant 
rate is zero and the superradiant rate is just twice the rate for an 
individual atom, is never reached. In a recent experimental realization 
of sub- and superradiance from two laser-trapped ions, DeVoe and Brewer 
\cite{DeVB96} measured the spontaneous emission rate of photons
as a function of the ion-ion distance in a laser trap of planar 
geometry which was strong enough to bring the ions (Ba$_{138}^+$)
to a distance of the order of 1$\mu$m of each other.

The two double quantum dots behave in analogy to the two atoms considered 
above. For a positive bias, 
$\varepsilon\!=\!\varepsilon_L\!-\!\varepsilon_R\!>\!0$, 
the state $\ket{L}$ can be identified with the excited state and $\ket{R}$
with the ground state. The inelastic rate $\nu$ with which $\ket{L}$ decays to 
$\ket{R}$ can be calculated with Fermi's Golden Rule,
\begin{equation}
\nu = \frac{8 \, \pi \, T_c^2}{\hbar \, (\varepsilon^2 \!+\! 4 T_c^2)} \; 
        \rho\big(\sqrt{\varepsilon^2 \!+\! 4 T_c^2}\,\big). 
\end{equation}
In contrast to a two level atom, a third state $\ket{0}$ exists in the 
double quantum dot as the additional electron can tunnel into the leads.

The use of triplet and singlet states as defined in Eq.~(\ref{basis_ts})
allows us to find an analytical result for the stationary current that 
quantitatively coincides with the exact numerical solution extremely well.
We consider the rate equation for the probabilities of the corresponding 
nine states and take into account the doubling of the inelastic rates due 
to the Dicke effect in the triplet channel, 
\begin{equation}
\label{bilanz_glg}
\begin{split}
\dot{p}_{00} &= \Gamma \, p_{R0} + \Gamma \, p_{0R} - 2 \Gamma \, p_{00},\\
\dot{p}_{L0} &= \Gamma \, p_{00} + \frac{1}{2} \, \Gamma \, p_{T_0}
             + \frac{1}{2} \, \Gamma \, p_{S_0} - (\nu + \Gamma) \,p_{L0},\\
\dot{p}_{0L} &= \Gamma \, p_{00} + \frac{1}{2}\, \Gamma \, p_{T_0} 
              + \frac{1}{2} \, \Gamma \, p_{S_0} - (\nu+\Gamma )\, p_{0L},\\
\dot{p}_{R0} &= \nu \, p_{L0} + \Gamma \, p_{T-} - 2 \Gamma \, p_{R0},\\
\dot{p}_{0R} &= \nu \, p_{0L} + \Gamma \, p_{T-} - 2 \Gamma \, p_{0R},\\
\dot{p}_{T+} &= \Gamma \, p_{0L} + \Gamma \, p_{L0} - 2 \nu \, p_{T+},\\
\dot{p}_{T_0} &= 2 \nu \, p_{T+} + \frac{1}{2}\, \Gamma \, p_{0R}
             +\frac{1}{2}\, \Gamma \, p_{R0} - (2 \nu +\Gamma) \, p_{T_0},\\
\dot{p}_{T-} &= 2 \nu \, p_{T_0} - 2 \Gamma \, p_{T-}, \\
\dot{p}_{S_0}&= \frac{1}{2}\, \Gamma \, p_{0R} + \frac{1}{2}\, \Gamma \, p_{R0}
               - \Gamma \, p_{S_0}.
\end{split}
\end{equation}

\begin{figure}
\centering
\psfrag{g}{$g$}
\psfrag{DI1}{\hspace*{-5mm}$\Delta I_1 /$pA}
\psfrag{T-S}{\hspace*{-5mm}$p_{T_0}-p_{S_0}$}
\psfrag{bilanz}{Rate equation~(\ref{bilanz_glg})}
\psfrag{num}{Master equation~(\ref{master_operator})}
\epsfig{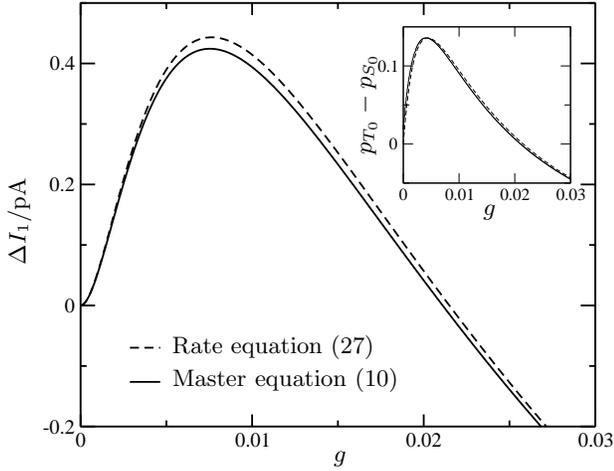}
\caption{
Enhancement of the tunnel current $\Delta I_1$ 
at the resonance $\varepsilon_1\!=\!\varepsilon_2\!=\!30\mu$eV
as a function of the dimensionless electron phonon coupling constant $g$,
Eq.~(\ref{ac_phonons}).
The additional current vanishes at $g\!\approx\!0.02$
when the tunnel rates to the double dot and between the dots become equal,
$\nu\!=\!\Gamma$.
The inset shows the difference in probabilities for triplet and singlet.}
\label{F_bilanz}
\end{figure}

Here, identical tunnel rates to all four leads have been assumed, 
$\Gamma_{L,1}\!=\!\Gamma_{R,1}\!=\!\Gamma_{L,2}\!=\!\Gamma_{R,2}\!=\!\Gamma$,
and $p_{L0}$ denotes the probability to find the first double dot in state 
$\ket{L}$ and the second in state $\ket{0}$.
Electrons can also tunnel into and out of the singlet state due to the 
coupling to the leads which is not possible in the original Dicke model.
In the stationary case, the Eq.~(\ref{bilanz_glg}) can be easily
solved. For the current through one of the two double dots we obtain
\begin{equation} \label{Idouble1}
\begin{split}
I_1 &= \frac{e\,\Gamma}{\hbar} \; (p_{00}+p_{0L}+p_{0R}) \\
    &=  \frac{e\,\Gamma}{\hbar} \; \frac{x (4x+1)}{9x^2+5x+1}, 
    \quad x=\nu / \Gamma.
\end{split}
\end{equation}
This can be compared with the tunnel current through one independent 
double dot, $I_1^0$ obtained by a similar rate equation,
\begin{equation}
I_1^0=\frac{e\,\Gamma}{\hbar} \; \frac{x}{1+2x}.
\end{equation}
The difference $\Delta I_1 = I_1 - I_1^0$ represents  the additional current 
due to the Dicke effect and is shown in Fig.~\ref{F_bilanz} as a function of 
the dimensionless coupling strength $g$ to the bosonic environment, together 
with a comparison to the $\Delta I_1$ as obtained from the numerical solution 
of Eq.~(\ref{master_operator}). Both results agree very well, indicating 
that it is indeed the Dicke effect that leads to the increase in the tunnel 
current. In addition, we show (inset of Fig.~\ref{F_bilanz}) 
the difference between triplet and singlet occupation probability
that follow from the Eq.~(\ref{bilanz_glg}) as 
\begin{equation}
p_{T_0} - p_{S_0} = - \frac{2x (x+2) (x-1)}{9x^3+23x^2+11x+2}.
\end{equation}
This is in excellent agreement with the numerical results 
and underlines that the change
in the tunnel current due to collective effects is proportional to
$p_{T_0} - p_{S_0}$, as already discussed above.
This demonstrates that the effect of superradiance amplifies
the tunneling of electrons from the left to the right dots
resulting in an enhanced current through the two double quantum dots.

\section{Current subradiance and inelastic switch}\label{sectionsub}
The close analogy with the Dicke effect suggests the existence of not only 
current super-, but also current subradiance in the register.
In the subradiant regime, the two DQDs form a singlet state where
the tunneling from the left to the right quantum dots is diminished,
resulting in a weaker tunnel current through the dots.

\subsection{Current antiresonance}
\begin{figure}
\centering
\psfrag{e1}{\hspace*{-4mm}$\varepsilon_1/\mu$eV}
\psfrag{I1}{\hspace*{-5mm}$I_1/$pA}
\psfrag{ee1}{\hspace*{-4mm}$\varepsilon_1/\mu$eV}
\psfrag{II1}{\hspace*{-10mm}
 $\big\langle \textrm{P}_{T_0} \big\rangle 
  - \big\langle \textrm{P}_{S_0}\big\rangle$}
\psfrag{Gr15}{$\Gamma_{R,2}= 0.15$}
\psfrag{Gr06}{$\Gamma_{R,2}= 0.06$}
\psfrag{Gr02}{$\Gamma_{R,2}= 0.02$}
\psfrag{Gr00}{$\Gamma_{R,2}= 0.00$}
\epsfig{file=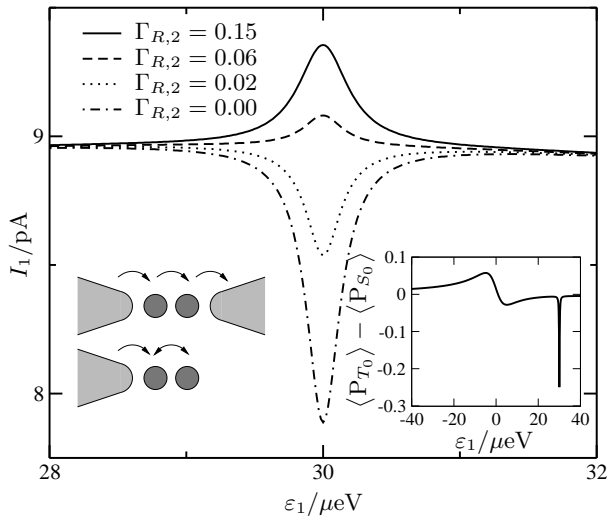,width=8cm}
\caption{Transition from an increased to a decreased
current through the first double quantum dot for different tunnel rates 
$\Gamma_{R,2}$ (in $\mu$eV) and  $\varepsilon_2\!=\!30\mu$eV.
The left inset shows schematically the set-up for $\Gamma_{R,2}\!=\!0$ and the
right inset gives the difference of triplet and singlet for the same  case.}
\label{F_sub}
\end{figure}

Subradiance occurs in our system in a slightly changed set-up where
electrons in the second double dot are prevented from tunneling into the 
right lead, $\Gamma_{R,2}\!=\!0$, as indicated in the
inset of Fig.~\ref{F_sub}. Then, the additional electron is trapped and
no current can flow through the second double dot.
Nevertheless, this electron can affect the tunnel current through the 
first double dot: Instead of a maximum, we now find a minimum at the 
resonance $\varepsilon_1\!=\!\varepsilon_2$.
Fig.~\ref{F_sub} shows how the positive peak in the current $I_1$
develops into a minimum as the tunneling rate $\Gamma_{R,2}$ is decreased
to zero. This minimum is indeed related to an increased probability of
finding the two dots in the singlet state $\ket{S_0}$ rather than in the
triplet state $\ket{T_0}$, as can be seen from the inset of Fig.~\ref{F_sub}.
Thus, in this regime the effect of subradiance dominates, 
leading to a decreased current.

This behavior is again consistent with the approximation Eq.~(\ref{rho6}) 
for the cross coherence $\big\langle p_1^{\dagger} p_2 \big\rangle$.
Taking into account the different non-interacting
matrix elements in the two double dots, 
$\erw{n_{L,1}}\!\neq\!\erw{n_{L,2}}$ and $\erw{p_1}\!\neq\!\erw{p_2}$ 
due to $\Gamma_{R,1}\!\neq\!\Gamma_{R,2}$, 
we find a negative cross coherence at the resonance from Eq.~(\ref{rho6}).
This corresponds to an increased probability for the singlet state
and according to Eq.~(\ref{Delta_I1}) to a negative peak in the tunnel current,
in agreement with our  numerical solution.

\subsection{Inelastic current switch}
Up to now, we have regarded the cross coherence
$\big\langle p_1^{\dagger} p_2 \big\rangle$
and its effects on the current only at the resonance 
$\varepsilon_1\!=\!\varepsilon_2$.
However, it was already pointed out in section~\ref{subsec_analytisch}
that another cross coherence, $\big\langle p_1 p_2 \big\rangle$,
exhibits a resonance if the bias in one dot equals the negative bias in the 
other dot, $\varepsilon_1\!=\!-\varepsilon_2$ (cp. Fig.~\ref{F_rho6}).
This case is considered in the following.

\begin{figure}[t]
\centering
\psfrag{e1}{\hspace*{-4mm}$\varepsilon_1/\mu$eV}
\psfrag{I2}{\hspace*{-3mm}$I_2/$pA}
\epsfig{file=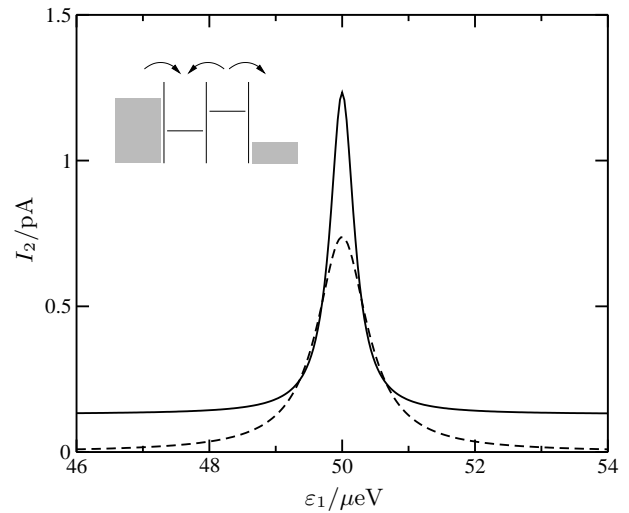,width=8cm}
\caption{Tunnel current through the second double dot which is blocked 
due to a negative bias, $\varepsilon_2\!=\!-50\mu$eV, 
as depicted in the inset ($g\!=\!0.015$).
The approximation for $\Delta I_2$, Eq.~(\ref{Delta_I2}) (dashed line),
agrees well with the result of the 
master equation~(\ref{master_operator}) (solid line -- the finite 
offset of which is the tail of the elastic current at $\varepsilon_2\!=\!0$).}
\label{F_block}
\end{figure}

We use a fixed negative bias $\varepsilon_2\!<\!0$ in the second double dot
as indicated in the inset of Fig.~\ref{F_block}. Consequently, electrons 
cannot tunnel from the left to the right dot such that the second double dot 
is blocked and no current can flow through it.
The presence of the first double dot, though, lifts this blockade and enables 
a current through the second double dot if the resonance condition
$\varepsilon_1\!=\!-\varepsilon_2$ is fulfilled. The  current $I_2$ is shown 
in Fig.~\ref{F_block} as a function of the bias in the first double dot,
$\varepsilon_1$.
Due to the coupling to the common phonon environment, energy is transferred
from the first to the second double dot, allowing electrons to tunnel
from the left to the right in the second double dot.
At the same time, the current through the first double dot is decreased 
(not shown here).

We can approximate the current through the second double dot around
$\varepsilon_1\!=\!-\varepsilon_2$ taking into account only 
$\big\langle p_1 p_2 \big\rangle$ in Eq.~(\ref{Delta_current}).
A similar calculation as for $\big\langle p_1^{\dagger} p_2 \big\rangle$,
Eq.~(\ref{rho6}), gives
\begin{equation}
\label{Delta_I2}
\begin{split}
\Delta I_2 = \frac{2 T_{c,2}\,\gamma_1 \,e}{\varepsilon_2 \,\hbar}\;
\frac{\frac{1}{2}\Gamma_{R,1}\!+\!\frac{1}{2}\Gamma_{R,2}\!+\!8 \alpha}
{(\varepsilon_1\!+\!\varepsilon_2)^2 
   + (\frac{1}{2}\Gamma_{R,1}\!+\!\frac{1}{2}\Gamma_{R,2}\!+\!8 \alpha)^2}\\
\Big(\gamma_1  \, \textrm{Re}\big\{\!\erw{p_2}\!\big\} \, \erw{n_{L,1}}
  + \gamma_2 \, \textrm{Re}\big\{\!\erw{p_1}\!\big\} \, \erw{n_{L,2}}\Big),
\end{split}
\end{equation}
with $\alpha\!=\!\alpha_1\!=\!\alpha_2$ evaluated at the resonance, 
where both systems are identical except of the bias.
This approximation again is in good agreement with the numerical solution of 
Eq.~(\ref{master_operator}), as can be seen from Fig.~\ref{F_block}.

Our results suggest that the current through one of the DQDs  can be 
switched on and off by appropriate manipulation of the other one.
We emphasize that this mechanism is mediated by the dissipative phonon 
environment and not the Coulomb interaction between the charges.
As this effect is very sensitive to the energy bias, it allows to detect
a certain energy bias in one double dot by observing the current through 
the other double dot. 

\section{Conclusion} \label{sectionconclusion}
In this work, we have investigated collective effects in two double quantum
dots. An indirect interaction arises between the two double dots due to the
coupling to the same phonon environment.
We predict that the Dicke effect causes a considerable increase or decrease
of the tunnel current, depending on the choice of the parameters.
The occurrence of the Dicke effect in the transport through mesoscopic 
systems has already been pointed out by Shahbazyan and Raikh \cite{SR94}.
In their system, the coupling to the same lead is responsible for collective
effects.
Usually, the Dicke effect manifests itself in a dynamic process like
the spontaneous emission of an ensemble of identical atoms 
\cite{DeVB96,Greiner_00}. Transport through double quantum dots, however, 
allows to study a time independent form of the Dicke effect.
Moreover, we have demonstrated that the change of the tunnel current
is connected with an entanglement of the different double dots.
This opens the possibility to realize and to measure specific entangled 
states of two double dots.
In particular, one can switch from a predominant triplet 
superposition of the two double dots connected with an increased tunnel current
to a predominate singlet state leading to a reduced current.

The results discussed here
were derived for the ideal case of an identical electron-phonon
coupling in both double quantum dots.
Furthermore, the Coulomb interaction between the two double dots
has not been considered here.
In a real experiment, these assumption will never be perfectly fulfilled
and would 
lead to deviations from the collective effects presented above. However, 
we predict that even  in presence of inter-dot Coulomb interactions, 
phonon mediated collective effects should persist as long as a description 
of the register in terms of few many-body states is possible. 
These many-body states (that would depend on the specific geometry of the 
register) would than replace the many-body basis 
$\{|0,i\rangle, |L,i\rangle,  |R,i\rangle\}$ ($i=1,2$) used in our model here.

We have derived the master equation for the general case of 
$N$ double dots but only focused on $N=2$ which is the simplest case where 
collective effects occur. In general, one of the main characteristic features 
of superradiance is the quadratic increase of the effect with increasing 
number of coupled systems. 
For the spontaneous collective emission from $N$ excited two
level atoms, this means that the maximum of the intensity of the emitted 
radiation increases with the square number of systems, $N^2$, while the time 
in which the decay takes place decreases inversely to the number of systems, 
$1/N$. 
Therefore, we expect that the collective effects as presented here
become even more pronounced if more than two double dots are indirectly coupled
by the common phonons.

\begin{acknowledgments}
We acknowledge B. Kramer for fruitful discussions.
This work was supported by projects EPSRC GR44690/01, DFG Br1528/4-1,
the WE Heraeus foundation and the UK Quantum Circuits Network.
\end{acknowledgments}

\appendix
\section{Master Equation for two double quantum dots}
\label{appendix_A}
The dimension of the density matrix $\rho$ for $N$ double quantum dots is equal
to $9^N$ such that the master equation~(\ref{master_operator}) corresponds
to 81 coupled differential equations for $N\!=\!2$.
It is, however, not necessary to solve all 81 equations as we study the current
which requires the knowledge of only six matrix elements, 
cp. Eq.~(\ref{current}). The smallest closed subset of equations, containing 
the equations for those six elements consists of 25 equations.

The mixed terms in the master equation~(\ref{master_operator}), $i\ne j$,
describing the indirect interaction between the two DQDs due to
the coupling to the same phonons, are marked in the following with an
additional prefactor $q$. Setting $q\!=\!0$ results in the master equation
for two completely independent double dots coupled to independent 
phonons. The interacting case corresponds to $q\!=\!1$.
Note that the elements of the density matrix are expressed with respect
to the basis $\{$$\ket{L}, \ket{R}, \ket{0}$$\}$ for each double dot.
Due to the tunneling of electrons between the left and right quantum dot,
these states are no eigenstates of the unperturbed Hamiltonian.
Finally, the master equation for the elements of the density matrix reads
\begin{widetext}
\begin{equation}
\label{master_elements}
\begin{split}
\dot{\rho}_{LLLL} &= i T_{c,1} (\rho_{LLRL}-\rho_{LRLL}) 
 + i T_{c,2} (\rho_{LLLR}-\rho_{RLLL}) 
 + \Gamma_{L,1} \, \rho_{L00L} +  \Gamma_{L,2} \, \rho_{0LL0}, \\ 
\dot{\rho}_{LLLR} &= i T_{c,1} (\rho_{LLRR}-\rho_{LRLR}) 
 + i  T_{c,2} (\rho_{LLLL}-\rho_{RLLR}) 
 + \Gamma_{L,1} \,\rho_{L00R} - \gamma_2 \,\rho_{LLLL} 
 - \beta_2  \,\rho_{RLLR} \\ &
 - \big(i \varepsilon_{2} + \frac{1}{2} \Gamma_{R,2} + 2 \alpha_2 \big)\,
   \rho_{LLLR} + q \, \beta_1 \big(\rho_{LLRR} - \rho_{LRLR} \big),\\
\dot{\rho}_{LLRL} &= 
 i T_{c,1} (\rho_{LLLL}-\rho_{LRRL}) + i T_{c,2} (\rho_{LLRR}-\rho_{RLRL})
 + \Gamma_{L,2}\,\rho_{0LR0} - \gamma_1 \,\rho_{LLLL}- \beta_1 \,\rho_{LRRL}\\&
 - \big(i \varepsilon_{1} + \frac{1}{2}\Gamma_{R,1} + 2\alpha_1 \big) \,
   \rho_{LLRL}
 + q \, \beta_{2} \big(\rho_{LLRR} - \rho_{RLRL}\big), \\
\dot{\rho}_{LLRR} &= i T_{c,1} (\rho_{LLLR}-\rho_{LRRR}) 
 + i T_{c,2} (\rho_{LLRL}-\rho_{RLRR})
 - \gamma_1 \,\rho_{LLLR} - \beta_1  \,\rho_{LRRR}\\&
 - \gamma_2 \,\rho_{LLRL}  - \beta_2  \,\rho_{RLRR}
 - \big(i \varepsilon_{1} + i \varepsilon_2 + \frac{1}{2}\Gamma_{R,1}
 + \frac{1}{2}\Gamma_{R,2} + 2 \alpha_1 + 2 \alpha_2 \big)\,\rho_{LLRR}\\&
 - q \;\big(\!2(\alpha_1\!+\!\alpha_2)\,\rho_{LLRR}
 + \beta_{2} \,\rho_{RLRR} + \beta_1\,\rho_{LRRR} + \gamma_2\,\rho_{LLRL}
 + \gamma_1 \,\rho_{LLLR}\big), \\
\dot{\rho}_{RLLR} &= i T_{c,1} (\rho_{RLRR}-\rho_{RRLR}) 
 + i T_{c,2} (\rho_{RLLL}-\rho_{LLLR})
 + \Gamma_{L,1}\,\rho_{R00R} - \Gamma_{R,2} \,\rho_{RLLR}, \\
\dot{\rho}_{RLRL} &= 
 i T_{c,1} (\rho_{RLLL}-\rho_{RRRL}) + i T_{c,2} (\rho_{RLRR}-\rho_{LLRL})
 - \gamma_1 \,\rho_{RLLL}  - \beta_1  \,\rho_{RRRL}\\&
 - \gamma_2 \,\rho_{LLRL}  - \beta_2 \,\rho_{RLRR}
 - \big(i \varepsilon_1 \!- i \varepsilon_{2} + \frac{1}{2}\Gamma_{R,1}
 + \frac{1}{2}\Gamma_{R,2} + 2 \alpha_1 + 2 \alpha_2 \big) \,\rho_{RLRL}\\&
 + q \;\big(2(\alpha_1\!+\!\alpha_2)\,\rho_{RLRL}
 + \gamma_2\,\rho_{LLRL} + \beta_{1} \,\rho_{RRRL} + \beta_2 \,\rho_{RLRR} 
 + \gamma_1 \,\rho_{RLLL}\big),\\
\dot{\rho}_{RLRR} &=
 i T_{c,1} (\rho_{RLLR}-\rho_{RRRR})+ i T_{c,2} (\rho_{RLRL}-\rho_{LLRR})
 - \gamma_1 \,\rho_{RLLR} - \beta_1  \,\rho_{RRRR} \\&
 - \big(i \varepsilon_{1} + \frac{1}{2}\Gamma_{R,1} + \Gamma_{R,2}
 + 2 \alpha_1 \big) \,\rho_{RLRR}
 + q \, \gamma_2 \big(\rho_{LLRR} - \rho_{RLRL}\big), \\
\dot{\rho}_{0LL0} &= i T_{c,1} (\rho_{0LR0}-\rho_{0RL0})
 + \Gamma_{L,1}\,\rho_{0000} + \Gamma_{R,2}\,\rho_{RLLR}
 - \Gamma_{L,2}\,\rho_{0LL0}, \\
\dot{\rho}_{0LR0} &=
 i T_{c,1} (\rho_{0LL0}-\rho_{0RR0}) + \Gamma_{R,2} \,\rho_{RLRR}
 - \gamma_1 \,\rho_{0LL0} - \beta_1  \,\rho_{0RR0}
 - \big(i \varepsilon_{1} + \frac{1}{2}\Gamma_{R,1} + \Gamma_{L,2} \,
 + 2 \alpha_1 \big) \,\rho_{0LR0} , \\
\dot{\rho}_{LRRL} &= i T_{c,1} (\rho_{LRLL}-\rho_{LLRL})
 + i T_{c,2} (\rho_{LRRR}-\rho_{RRRL})
 + \Gamma_{L,2} \,\rho_{0RR0} - \Gamma_{R,1} \,\rho_{LRRL}, \\
\dot{\rho}_{LRRR} &= i T_{c,1} (\rho_{LRLR}-\rho_{LLRR})
 + i T_{c,2} (\rho_{LRRL}-\rho_{RRRR})
 - \gamma_2 \,\rho_{LRRL} - \beta_2  \,\rho_{RRRR}\\&
 - \big(i \varepsilon_{2} + \Gamma_{R,1} + \frac{1}{2}\Gamma_{R,2}
 + 2 \alpha_2 \big) \, \rho_{LRRR}
 + q \, \gamma_{1} \big(\rho_{LLRR} - \rho_{LRLR} \big), \\
\dot{\rho}_{RRRR} &= i T_{c,1} (\rho_{RRLR}-\rho_{RLRR})
 + i T_{c,2} (\rho_{RRRL}-\rho_{LRRR})
 - \big(\Gamma_{R,1} + \Gamma_{R,2}\big) \,\rho_{RRRR}, \\
\dot{\rho}_{0RR0} &= i T_{c,1} (\rho_{0RL0}-\rho_{0LR0})
 + \Gamma_{R,2} \,\rho_{RRRR}
 - \big(\Gamma_{R,1} + \Gamma_{L,2}\big) \,\rho_{0RR0}, \\
\dot{\rho}_{L00L} &= i T_{c,2} (\rho_{L00R}-\rho_{R00L})
 + \Gamma_{R,1} \,\rho_{LRRL}
 + \Gamma_{L,2} \,\rho_{0000} - \Gamma_{L,1} \,\rho_{L00L} , \\
\dot{\rho}_{L00R} &= 
 i T_{c,2} (\rho_{L00L}-\rho_{R00R}) + \Gamma_{R,1} \,\rho_{LRRR} 
 - \gamma_2 \,\rho_{L00L} - \beta_2  \,\rho_{R00R}
 - \big( i \varepsilon_{2} + \Gamma_{L,1} + \frac{1}{2} \Gamma_{R,2}
 + 2\alpha_2 \big) \,\rho_{L00R},\\
\dot{\rho}_{R00R} &= i T_{c,2} (\rho_{R00L}-\rho_{L00R})
 - \Gamma_{L,1} \,\rho_{R00R} + \Gamma_{R,1} \,\rho_{RRRR}
 - \Gamma_{R,2} \,\rho_{R00R},\\
\dot{\rho}_{0000} &= \Gamma_{R,1} \,\rho_{0RR0} + \Gamma_{R,2} \,\rho_{R00R}
 - \big(\Gamma_{L,1} + \Gamma_{L,2}\big) \,\rho_{0000}.
\end{split}
\end{equation}
\end{widetext}
The remaining 8 equations follow immediately since $\rho$ is an hermitian
operator,
\begin{equation}
\rho_{j\,i\,i'\,j'} = \rho_{j'\,i'\,i\,j}^*\,,
\end{equation}
and the coefficients $\alpha_j$, $\beta_j$, and $\gamma_j$ are defined as
\begin{equation}
\label{rate_abc}
\begin{split}
\alpha_j &= \frac{4 \pi T_{c,j}^2}{\Delta_j^2} \; \rho(\Delta_j)\,
               \coth \!\Big(\frac{\beta \Delta_j}{2}\Big),\\
\beta_j  &= \frac{2 \pi T_{c,j}}{\Delta_j} \; \rho(\Delta_j) \,
     \Big( 1- \frac{\varepsilon_j}{\Delta_j}\, 
     \coth \!\Big(\frac{\beta \Delta_j}{2}\Big)\Big) ,\\
\gamma_j &= \frac{2 \pi T_{c,j}}{\Delta_j} \; \rho(\Delta_j) \,
     \Big( 1+ \frac{\varepsilon_j}{\Delta_j}\, 
     \coth \!\Big(\frac{\beta \Delta_j}{2}\Big)\Big).
\end{split}
\end{equation}


\end{document}